\documentstyle[twoside,fleqn,epsfig]{actaps}

\begin{document}

\headings{1}{6}
\def\authorlist{K. Banaszek}
\def\shorttitle{Maximum-likelihood algorithm for quantum tomography}

\title{MAXIMUM-LIKELIHOOD ALGORITHM FOR QUANTUM TOMOGRAPHY}
\author{{Konrad Banaszek}\email{Konrad.Banaszek@fuw.edu.pl}}
{Instytut Fizyki Teoretycznej, Uniwersytet Warszawski, Ho\.{z}a~69,
PL--00--681~Warszawa, Poland\\
and Istituto Nazionale di Fisica della Materia -- Unit\'{a} di Pavia\\
via A. Bassi 6, I-27100 Pavia, Italy}

\day{15 April 1999}

\abstract{Optical homodyne tomography is discussed in the context of
classical image processing. Analogies between these two fields are traced
and used to formulate an iterative numerical algorithm for reconstructing
the Wigner function from homodyne statistics.}

\medskip


\section{Introduction}
Several years after the first demonstration \cite{R1}, optical homodyne
tomography has become a well established tool for measuring quantum
statistical properties of optical radiation. What is particularly
fascinating, this technique provides practical means to visualise the
measured quantum state in the form of the Wigner function. This success
is a result of combining a complete quantum mechanical measurement of
field quadratures with a filtered back-projection algorithm used in
medical imaging.

The purpose of this contribution is to trace some other analogies
between quantum state reconstruction and classical image processing,
with the motivation to develop novel numerical methods for quantum
tomography. Our interest will be focused on imperfect detection
\cite{R2,R3}, which has deleterious effects on quantum interference phenomena
exhibited by non-classical states \cite{R4}. As we will discuss in Sec.~2,
such effects can be related in the phase space representation to image
blurring. Restoration of blurred images is a well known problem in
classical imaging, and a number of methods has been developed for this
purpose. Specifically, we shall briefly describe in Sec.~3 the Richardson
algorithm \cite{R5} (known also in statistics as the expectation-maximization
algorithm \cite{R6}), which provides an effective iterative procedure to
perform image deblurring.

An interesting question is, whether classical deblurring methods can
be transferred to quantum tomography. We discuss this, in the case of
the Richardson algorithm, in Sec.~4. The answer is not straightforward:
the Richardson algorithm assumes positive definiteness of the original
undegraded image, and this condition is not satisfied by the quantum
mechanical Wigner function. We will show that this difficulty can be
overcome by expressing the Wigner function in terms of the phase space
displaced photon distribution. This yields an iterative algorithm for
reconstructing the Wigner function, which incorporates compensation for
detection losses in a numerically stable way \cite{R7}.

\section{Imperfect detection and image blurring}
Homodyne detection is a realization of the quantum mechanical measurement
of field quadratures only in the idealized loss-free limit.
In practice, a fraction of the signal field is always lost
due to the mode-mismatch and the non-unit efficiency of photodiodes.  
The homodyne statistics collected using
a realistic setup is described by the distribution \cite{R8}:
\begin{equation}
\label{Eq:hxtheta}
h(x;\theta) =
\int_{-\infty}^{\infty} \mbox{d}x' \;
\frac{1}{\sqrt{\pi(1-\eta)}} \exp \left(
- \frac{(x-\sqrt{\eta}x')^{2}}{1-\eta}
\right)
\langle x'_{\theta} | \hat{\rho} | x'_{\theta} \rangle,
\end{equation}
where $\theta$ is the local oscillator phase, $\eta$ characterizes
the overall detector efficiency, and $\langle x'_{\theta} | \hat{\rho}
| x'_{\theta} \rangle$ denote diagonal elements of the density matrix
$\hat{\rho}$ in the quadrature basis.

Application of the back-projection transformation to $h(x;\theta)$
yields, instead of the Wigner function, a generalized phase space
quasidistribution function $P(q,p; s)$ with the ordering parameter
$s=-(1-\eta)/\eta$. This function can be expressed as a
convolution of the Wigner function $W(q,p)$ with a gaussian function:
\begin{equation}
\label{Eq:Palphas}
P(q,p; -|s|) = \int \mbox{d}q'\,\mbox{d}p' \; \frac{1}{\pi |s|}
\exp\left( - \frac{1}{|s|}[(q-q')^2+(p-p')^2 ]\right) W(q',p').
\end{equation}
Thus what we reconstruct from imperfect homodyne data,
is a blurred version of the Wigner function. The question is, whether we
can get rid of this blurring in numerical processing of experimental data.

A similar problem appears in the following classical context:
suppose we observe an image using an imperfect apparatus (for example
ill-matched glasses), which generates some blurring. Such blurring can be
described by a so-called point spread function specifying the shape
generated by a single point of the original image. The observed
degraded image is consequently
given by a convolution of the original image with
the point spread function. Using this language, we can assign the 
following names to the terms of Eq.~(\ref{Eq:Palphas}):
\begin{center}
\begin{tabular}{rcl}
$W(q',p')$ & --- & original image \\
$ \displaystyle
\frac{1}{\pi |s|} \exp\left( - \frac{1}{|s|} 
[(q-q')^2+(p-p')^2]\right)$
& --- & point spread function \\
$P(q,p; -|s|)$ & --- & degraded image
\end{tabular}
\end{center}
The common problem now is the restoration of the original image
from the degraded one, assuming that the point spread function
is known.

\section{Image restoration}
An analytical way to deconvolve Eq.~(\ref{Eq:Palphas}) is to apply
the Fourier transform, which maps a convolution onto a direct product.
Dividing both the sides by the Fourier-transformed point spread function
and evaluating the inverse Fourier transform thus yields an explicit
expression for the original image.  However, this procedure is very
sensitive to statistical fluctuations and numerical truncation errors,
which makes its practical application a very delicate matter. These 
problems have been noted also in the context of quantum tomography \cite{R9}.

The numerical instability of the Fourier deconvolution has led to the
development of techniques dedicated for image restoration. The
basic observation is that statistical noise does not allow us to specify
precisely the original image that was 'behind' the blurred data. In
principle, the measured degraded image could could be generated by a
variety of original images. However, comparing various original images we
intuitively expect that some of them were {\em more likely} to generate
the measured data than other ones. The maximum-likelihood methodology
quantifies this intuition, and selects as an estimate the original image
which maximizes the likelihood. 

In order to discuss this idea in detail, we shall
consider a discretized version of Eq.~(\ref{Eq:Palphas}):
\begin{equation}
\label{Eq:LININPOS}
p_{\nu} = \sum_{n} A_{\nu n} w_{n},
\end{equation}
where $w_n$ is the original image,
$A_{\nu n}$ is the  point spread function, and
$p_{\nu}$ is the degraded image. Note that this formulation
is more general compared to Eq.~(\ref{Eq:Palphas}), because
it allows the point spread function to be of different form
for each 'element' of the original image indexed with $n$.
The likelihood can be quantified using the function
\begin{equation}
{\cal L} = \sum_{\nu} p_{\nu} \ln \left(
\sum_{n} A_{\nu n} w_{n} \right)
\end{equation}
which has a rigorous derivation when the degraded image is observed as a
histogram of events governed by Poissonian statistics.  The likelihood
function for quantum measurement has been discussed in Ref.~\cite{R10},
where its maximization has been proposed as a method for quantum state
estimation.

In classical imaging, it is natural to assume that $w_{n}$, as well as
$A_{\nu n}$ as a function of $\nu$ for each $n$, are positive definite
distributions with sum equal to one. 
Under these assumptions, it is
possible to find the maximum of the likelihood function ${\cal L}$
via simple iteration:
\begin{equation}
w_{n}^{(i+1)} = \sum_{\nu} p_{\nu} 
\frac{A_{\nu n} w_{n}^{(i)}}{\sum_{m} A_{\nu m} w_{m}^{(i)}},
\end{equation}
which is the essence of the Richardson algorithm for image
restoration~\cite{R5}. A simple heuristic derivation of this algorithm
can be found in Ref.~\cite{R7}.

Of course, the maximum-likelihood approach is not a magic wand solving
unconditionally the problem of image restoration. With increasing
blurring, the quality of the reconstructed image worsens, and the
convergence of the iterative algorithm may be very slow. In many cases,
however, it offers superior performance compared to the Fourier
deconvolution technique.

\section{Quantum tomography}
An obvious difficulty with applying the iterative restoration algorithm to
quantum tomography is that the object to be reconstructed in the quantum
case, i.e.\ the Wigner function, is not positive definite.  Nevertheless,
there are some other quantum mechanical reconstruction problems, where
positivity constraints appear in a natural way.  An interesting and
nontrivial example is determination of the photon number distribution
from random phase homodyne statistics \cite{R11}. The relation between the
phase-averaged homodyne statistics and the photon number distribution
$\varrho_n$ is given by
\begin{equation}
\frac{1}{2\pi}\int_{-\pi}^{\pi} \mbox{d}\theta \;
h(x;\theta) = \sum_{n} A_n(x) \varrho_n,
\end{equation}
where $A_{n}(x)$ describe contributions to the homodyne statistics
generated by different Fock states $|n\rangle$.
This formula, after discretization of $x$, is exactly of the form
assumed in Eq.~(\ref{Eq:LININPOS}).
Thus we arrive at the following formal analogy:
\begin{center}
\begin{tabular}{rcl}
$\varrho_n$ & --- & original image \\
$A_{n}(x)$ & --- & point spread function \\
$\displaystyle \frac{1}{2\pi} 
\int_{-\pi}^{\pi}
\mbox{d}\theta \; h(x; \theta)$ & --- &
degraded image
\end{tabular}
\end{center}
which allows us to apply directly the iterative reconstruction algorithm
\cite{R12}. In this procedure, there is one {\em a priori} parameter: it is the
cut-off of the distribution $\varrho_n$ specifying the maximum number
of photons.

Reconstruction of the photon distribution may seem to be quite distant
from the starting point of our considerations, which was deblurring
of the Wigner function. However, let us recall that the Wigner function
can be represented as an alternating sum of the photon distribution
$\varrho_{n}(q,p)$ corresponding to the phase space displaced state
$\hat{D}^{\dagger}(q,p)\hat{\varrho}\hat{D}(q,p)$:
\begin{equation}
\label{Eq:Wqp}
W(q,p) = \frac{1}{\pi} \sum_{n=0}^{\infty} (-1)^{n}
\varrho_{n}(q,p)
\end{equation}
Obviously, we can apply this formula to evaluate the Wigner function
at $q=p=0$. What would be of interest, is the generalization of the
maximum-likelihood algorithm to determination of an arbitrarily displaced
photon distribution $\varrho_n(q,p)$. This would yield a numerically
stable procedure for reconstructing the Wigner function from homodyne
statistics, even in the case of the non-unit detection efficiency.

Surprisingly, this generalization is quite straightforward. The basic
observation is that the displacement transformation has a simple effect
on the homodyne statistics, shifting it by
$\sqrt{\eta}(q\cos\theta+p\sin\theta)$ for a given local oscillator
phase $\theta$. Consequently, we have the relation
\begin{equation}
\label{Eq:tomolininpos}
\frac{1}{2\pi}\int_{-\infty}^{\infty} \mbox{d}\theta \;
h(x+\sqrt{\eta} q \cos\theta + \sqrt{\eta}p \sin\theta ; \theta)
= \sum_{n} A_{n}(x) \varrho_{n}(q,p),
\end{equation}
which can be readily implemented in the iterative algorithm.
Thus, we have arrived at the following two-step algorithm for quantum
tomography: for a given phase space point $(q,p)$, construct the
phase-averaged histogram according to the right-hand side
of Eq.~(\ref{Eq:tomolininpos}), and iteratively reconstruct
$\varrho_n(q,p)$.  Then, calculate the value of the Wigner function
according to Eq.~(\ref{Eq:Wqp}). In Fig.~1 we present Monte Carlo
simulated reconstruction of
the Wigner function for a Schr\"{o}dinger cat state detected using
a homodyne setup with the efficiency $\eta=90\%$.

\begin{figure}
\centerline{\setlength{\unitlength}{1pt}
\begin{picture}(325,250)
\put(0,0){\epsfig{file=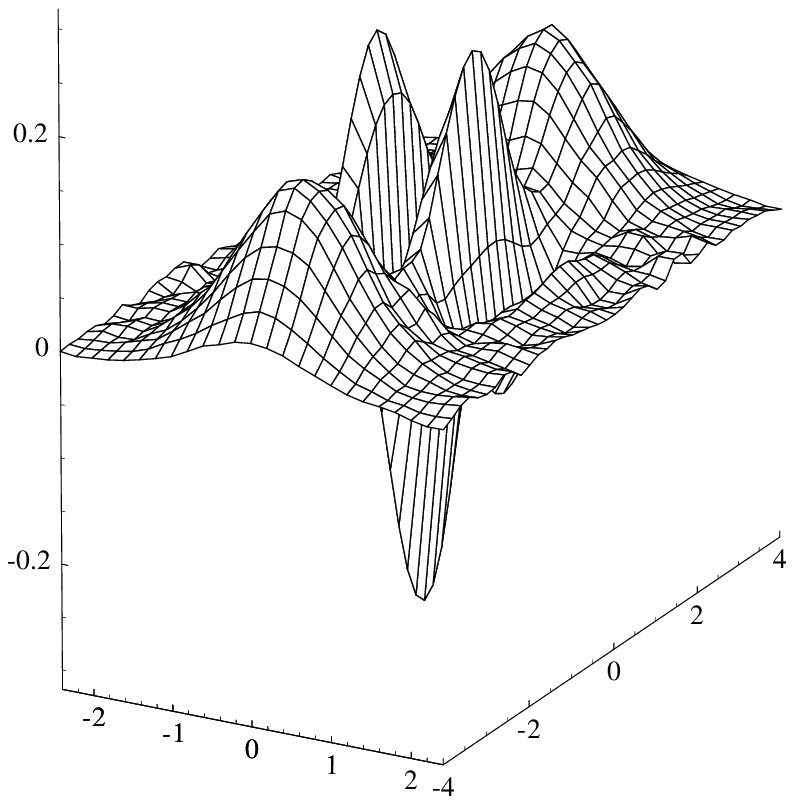}}
\put(20,225){\small $W(q,p)$}
\put(120,0){\small $q$}
\put(235,25){\small $p$}
\end{picture}}

\bigskip

\caption{Fig.~1. Reconstruction of the Schr\"{o}dinger cat state
$|\Psi\rangle\propto|2i\rangle -|-2i\rangle$ from Monte Carlo simulated
homodyne experiment. The homodyne data consisted of $10^5$ events
generated for each of $64$ phases spaced uniformly between $0$ and
$\pi$. At each point of the grid, the displaced photon statistics
was obtained from $10^4$ iterations, starting from a flat distribution
for $0\le n \le 39$. In the simulations, the homodyne variable $x$ has 
been discretized into 16000 bins over the range $-8\le x \le 8$.}
\end{figure}

The standard filtered back-projection algorithm used in quantum
tomography is based on the inverse Radon transform, whose integral
kernel is singular. Therefore, a regularization scheme is necessary
in processing experimental data. This aspect has a counterpart in the
maximum-likelihood algorithm. In this approach, we have the cut-off
for the photon distribution which can be regarded as a regularization
parameter. Its proper choice is an important matter. Setting it too
small perturbes the reconstructed photon distribution. On the other hand,
the larger number of $\varrho_n$s, the slower iterations converge. The
expected shape of the photon distribution  can be quite easily predicted,
if we roughly know the region of the phase space where the Wigner function
is localized. For this purpose it is useful to recall the semiclassical
picture of projections on Fock states as rings in the phase space
characterized by the radius $\sqrt{2n}$. The photon distribution is
nonzero over the range of $n$ for which the corresponding rings overlap
with the localization region for the Wigner function.

Truncation of the photon distribution can be introduced as a
regularization scheme also in the standard linear reconstruction
approach. In such a scheme, the Wigner function would be evaluated
from a finite part of the photon distribution reconstructed using the
pattern function technique. However, properties of the reconstructed
photon distribution make this method very sensitive to statistical
noise. This can be straightforwardly seen in the most regular case of
$\eta=1$. For large $n$, the error of $\varrho_n$ tends to a fixed nonzero
value \cite{R13}, and moreover deviations of consecutive $\varrho_n$s are
strongly anticorrelated. Consequently, the alternating sum defined in
Eq.~(\ref{Eq:Wqp}) accumulates the statistical uncertainty of the photon
distribution \cite{R14}.

Let us also note that in principle we could apply the restoration
algorithm to homodyne histograms described by Eq.~(\ref{Eq:hxtheta}),
in order to obtain deblurred quadrature distributions $\langle x_\theta
| \hat{\varrho} | x_\theta \rangle$. In this case statistical
fluctuations would play a much more significant role. The advantage of
using Eq.~(\ref{Eq:tomolininpos}) is that we use the whole available
sample of experimental data to determine a relatively small number of
parameters $\varrho_{n}(q,p)$.

\medskip

\noindent{\bf Acknowledgements}
I would like to thank Prof.\ G.~Mauro D'Ariano for his hospitality
during my stay in Pavia supported by INFM, and Prof.\ Krzysztof
W\'{o}dkiewicz for comments on the manuscript. I have benefited a lot
from discussions with Zdenek Hradil on maximum-likelihood methods in
quantum state reconstruction. This research is supported by
Komitet Bada\'{n} Naukowych, grant 2P03B~013~15.

\end{document}